\documentclass[aps,superscriptaddress]{revtex4-2}
\usepackage[T1]{fontenc}
\usepackage[utf8]{inputenc}
\usepackage{mathtools}
\usepackage{graphicx}
\usepackage{caption}
\usepackage{subcaption}
\usepackage{xcolor}
\usepackage{soul}
\usepackage{amsmath,amssymb}
\usepackage[colorlinks=true, pdfstartview=FitV, linkcolor=blue, citecolor=blue, urlcolor=magenta, breaklinks=true]{hyperref}
\usepackage{orcidlink}

\begin{document}

\author{Adamu Issifu \orcidlink{0000-0002-2843-835X}} 
\email{ai@academico.ufpb.br}

\affiliation{Instituto Tecnológico de Aeronáutica,\\ CEP 12.228-900, São José dos Campos, SP, Brazil} 

\author{Franciele M. da Silva \orcidlink{0000-0003-2568-2901}} 
\email{franciele.m.s@ufsc.br}

\affiliation{Departamento de F\'isica, CFM - Universidade Federal de Santa Catarina; \\ C.P. 476, CEP 88.040-900, Florian\'opolis, SC, Brazil.}

\author{D\'ebora P. Menezes \orcidlink{0000-0003-0730-6689}}
\email{debora.p.m@ufsc.br}

\affiliation{Departamento de F\'isica, CFM - Universidade Federal de Santa Catarina; \\ C.P. 476, CEP 88.040-900, Florian\'opolis, SC, Brazil.}

\title{Proto-strange quark stars from density-dependent quark mass model}

\begin{abstract}

In this paper, we investigate the evolution of strange quark stars (SQS) from birth as proto-strange quark stars to maturity as stable SQSs at a zero temperature. We assume that self-bound free quarks form {entirely the compact} star and study its evolution {through a series of snapshots}  using a density-dependent quark mass model. We consider $\beta$-equilibrated stellar matter at two major stages of the star's evolution: neutrino trapped regime and neutrino transparent regime during the deleptonization and cooling processes of the star. We fix the entropy density per baryon and the lepton fraction to investigate the nuclear equation of state (EoS), particle distribution, temperature profile inside the star, sound velocity, polytropic index, and the structure of the star. Our results show that stars with higher neutrino concentrations are slightly more massive than the neutrino-poor ones along the evolution lines of the SQS. We obtain EoSs in agreement with the conformal boundary set through sound velocity, and also the 2 M$_\odot$ mass constraint for NSs was satisfied at all stages of the star's evolution.

\end{abstract}

\maketitle

\section{Introduction}

Studying compact objects is an important fundamental research subject in contemporary nuclear physics, astrophysics, and cosmology. It plays a crucial part in understanding the early universe's nuclear structure, astrophysics, and state of matter. Compact objects enable us to explore the behavior of particles under exotic conditions such as higher baryon densities and temperatures in their core. The SQSs are compact objects whose existence was hypothesized for the first time by Ivanenko and Kurdgelaidze in 1965 \cite{Ivanenko:1965dg, Ivanenko:1969bs}. Witten later suggested that the fundamental state of hadronic matter could be a strange matter (SM) \cite{PhysRevD.30.272}. Nowadays, it has become a common phenomenon to explore the inner core of NSs assuming that it is composed of deconfined quark matter taking into account the clear mass-radius window imposed by the data from the Neutron Star Interior Composition Explorer (NICER) \cite{NANOGrav:2019jur, Fonseca:2021wxt, Riley:2021pdl} and the gravitational wave laser interferometer (Advanced LIGO, Virgo, and KAGRA) \cite{LIGOScientific:2017vwq, LIGOScientific:2016aoc, LIGOScientific:2018cki, LIGOScientific:2018hze, Miller:2023qyw}. Besides, the results found in \cite{Annala:2019puf} indicate that to satisfy the conformal bound for speed of sound, the massive NSs with $M \geq 2$ M$_\odot$ are expected to have a deconfined quark core. At higher baryon densities (several times the nuclear saturation density $n_0 = 0.152\,{\rm fm}^{-3}$) in the core of the compact objects, the particles are tightly packed overlapping each other such that they are capable of converting into free quarks. Therefore, two different forms of stars are expected: a hybrid NS (an NS with free quarks in its core) \cite{Issifu:2023ovi, Lopes:2021jpm} or a pure SQS \cite{dasilva2023bayesian}. %
Generally, quark stars (QS) are composed of high up ($u$) and down ($d$) quark asymmetry and can be formed entirely by deconfined SQM. A stable SQS can be formed from $u,\; d$ and $s$ (strange) quarks with leptons at $\beta$-equilibrium \cite{PhysRevD.4.1601, PhysRevD.30.272, Bombaci:2004mt, PhysRevD.84.083002, dasilva2023bayesian}.  

Proto-neutron stars (PNS) or black holes are formed after the gravitational collapse in the core of a massive star after it has exhausted its fuel supply depending on the initial conditions, triggering a type II supernova explosion. The PNSs at the stage of collapse trap neutrinos and become neutrino-rich objects. The star then goes through several stages of evolution after the core bounce, including heat transfer, neutrino diffusion, deleptonization, and entropy variation until the stellar matter becomes neutrino-poor, catalyzes, and starts cooling to form cold NSs several years after its birth \cite{Prakash:1996xs, Prakash:2000jr, Glendenning:1997wn}. A study of PNSs with exotic baryons in their core can be found in  \cite{Issifu:2023qyi, Raduta:2020fdn, Oertel:2016xsn, Sedrakian:2022kgj} while PNSs with free quarks in their core can be found in \cite{PhysRevC.100.015803, Shao:2011nu}. There is still limited information regarding the transition from PNSs to proto-strange quark stars (PSS) during supernova explosion due to the difficulty involved in the deconfinement simulation
process from hadronic matter to SQM \cite{Prakash:1996xs, Bombaci:2004mt, Graeff:2018czm}. {That notwithstanding, in this work we assume that the entire supernovae remnant (PNS) is composed of SQM. In this case, we do not expect any phase transitions since PSSs are assumed right from birth using thermodynamic conditions characteristic of PNSs evolution \cite{janka2012core, Fischer:2011zj} to investigate their evolution.}
Since SQM has been hypothesized to be the true ground state of quantum chromodynamics (QCD) \cite{Ivanenko:1969gs, 10.1143/PTP.44.291, PhysRevD.4.1601, PhysRevD.30.272}, hadronic matter can transition into QM, at this stage, there might be only QSs instead of NSs \cite{alcock1986strange}. It has been shown in Ref.\cite{alcock1986strange, Gupta:2002hj, shen2005slowly, Dexheimer:2013eua, Dexheimer:2012mk, Drago:2013fsa, drago2016scenario, bauswein2016exploring} that NS merger events can also lead to the formation of PSSs, which may be bare or have a dynamically irrelevant crust of nucleonic matter, although there are arguments against the existence of a crust in SQS \cite{kettner1995structure,usov1997low,melrose2006pair}.

The observation of massive compact objects sets a strict constraint on the EoS of the stellar matter, discounting most of the conventional phenomenological quark models that are not capable of producing heavy quark stars with $2\,\,\text{M}_\odot$ threshold. Additionally, the observation of binary merger GW190814 event \cite{abbott2016ligo} by LIGO/Virgo Collaborations, involving a secondary compact object with mass determined to be around $2.50-2.67\,\text{M}_\odot$ at $90\%$ confidence level has attracted extensive attention leading to further strict constraints. The observation of the first authoritative binary NS merger event with the emission of gravitational wave signal GW170817 by LIGO/Virgo Collaborations \cite{LIGOScientific:2017vwq, LIGOScientific:2018hze} reinforces these constraints. A recent data from NICER \cite{NANOGrav:2019jur, Fonseca:2021wxt} has led to the measurement of a massive pulsar PSR J0740+6620 with mass $2.072_{-0.066}^{+0.067}$ M$_{\odot}$ and radii $12.39^{+1.30}_{-0.98}\,{\rm km}$ at $68.30\%$ confidence level \cite{Riley:2021pdl} introducing a clear mass and radii window within which NSs can be explored. Besides, some compact stars are candidates for even more massive neutron stars, such as PSR J2215+5135 with a mass of $2.27_{-0.15}^{+0.17}\,\text{M}_\odot$ \cite{linares2018peering} and PSR J0952-0607 with a mass of $2.35 \pm 0.17$ M$_{\odot}$ \cite{romani2022psr}.

In this work, we explore newly born proto-strange stars formed during the gravitational collapse of a massive star as a neutrino-rich object to the formation of a stable neutrino transparent SQS at a zero temperature ($T=0$ MeV). {We assume that PNSs are composed entirely of self-bound quarks in a deconfined state and study their evolution as PSSs.} We employ a density-dependent quark mass model (DDQM) to calculate the EoS, particle abundances ($Y_i$), the structure of the star, temperature profile, polytropic index ($\gamma$), and speed of sound ($c_s$). We determine how the particle distribution inside the star during its evolution influences the softness or the stiffness of the EoS and the other NS characteristics during the early stages of the star's life. The DDQM involves two free model parameters $D$ (which represents the linear confining properties of the QM) and $C$ (which is related to the leading order perturbative behavior of the QM) that require fixing in the model framework. Different authors have adopted different approaches to determine suitable values of $D$ and $C$ that produce stable QSs with a maximum mass within the 2M$_\odot$ constraint
\cite{Backes:2020fyw, Wen:2005uf, Chen:2021fdj, Chen:2023rza}. However, in this work, we use $C = 0.7$ (dimensionless) and $\sqrt{D}=130.6$ MeV motivated by the results in \cite{dasilva2023bayesian} obtained using Bayesian analysis. This choice of parameters produces QSs that satisfy the mass-radius constraints imposed by PSR J0740+6620 \cite{Riley:2021pdl} and PSR J0030+0451 \cite{Riley:2019yda} pulsars. It also leads to a stable SQS at $T = 0$ MeV based on the Bodmer and Witten conjecture \cite{PhysRevD.30.272, PhysRevD.4.1601, Glendenning:1997wn} which hypothesized that energy density ($\varepsilon$) per baryon density ($n_b$) of SQM at zero pressure ($P=0$) should be smaller than the energy per nucleon ($E/A$) of $^{56}$Fe, \textit{i.e.}, $\left({\varepsilon}/{n_b}\right)_{\rm SQM}\leq 930\,{\rm MeV}$.
In the same vein, the two-flavor quark system should have $\varepsilon/n_b$ greater than the $E/A$ of $^{56}$Fe, $ \left({\varepsilon}/{n_b}\right)_{\rm 2QM} > 930\,{\rm MeV}$, otherwise protons and neutrons would decay into free $u$ and $d$ quarks. Also, we ensure that the fixed entropy SQSs determined within the model framework satisfy the $2\,\text{M}_\odot$ mass window required for NSs. Some other studies have also been done regarding PSSs using other quark models with different focuses on NS characteristics to describe the evolution of the SQSs in Ref.\cite{chu2021quark, Bordbar:2020fqj, Chu:2017huf}.

The paper is organized as follows: In Sec.~\ref{secdd} we present the general overview of the DDQM and elaborate on the zero temperature formalism in Subsec.~\ref{secdd1} and the finite temperature formalism in Subsec.~\ref{secdd2}. In Sec.~\ref{secprop} we discuss the properties of the SQM and the required conditions for calculating the EoS of the SQSs. We present our findings and discuss them in detail in Sec.~\ref{results} and our concluding remarks in Sec.~\ref{conc}.

\section{The DDQM Model} \label{secdd}

It is common knowledge that particle masses vary relative to the media in which they are measured. Such media-dependent masses are the so-called effective masses. In general, the effective masses vary, but the coupling constants also run in such media. Models in which the particle masses depend on chemical potential and/or temperature are referred to as the quasiparticle models \cite{PhysRevD.103.103021, goloviznin1993refractive}. Models of these kinds have been extensively studied in the literature \cite{Schertler:1996tq, Wen:2005uf, Backes:2020fyw, PhysRevC.61.045203, Gardim:2009mt, chu2021quark}. 

The original intention of the DDQM model was to use the density-dependent quark mass model to study the nonperturbative characteristics of the QM \cite{Fowler:1981rp, Plumer:1984aw}. It was later extended to calculate the EoS \cite{CHAKRABARTY1989112, PhysRevD.51.1989, PhysRevLett.74.1276, Peng:2000ff}, viscosity of SQM  and dissipation of r-modes \cite{PhysRevC.70.015803}, diquark properties \cite{Lugones:2002vd}, and compact astrophysical objects \cite{dasilva2023bayesian, Backes:2020fyw, chu2021quark} among others. The two main problems associated with the DDQM are the proper determination of the mass scaling and the thermodynamic consistency of density-dependent particle masses. These problems have widely been addressed in Refs.\cite{Peng:1999gh, Peng:2000ff, Chen:2021fdj, Backes:2020fyw} and references therein. In this work, we consider the case in which strong interquark interactions are caused by density and temperature-dependent quark masses. We investigate the evolution of the PSSs that require both fixed entropy density per baryon and $T=0$ MeV applications of the model, so we highlight each of these cases briefly below.

\subsection{Zero temperature formalism}\label{secdd1}

Considering a three-flavor quark ({ $u$, $d$, and $s$)} model at $T=0$ MeV with a finite chemical potential, $\mu_i$. The Fermi momentum $\nu_i$ in phase space is an important parameter, unlike the finite temperature treatment in the next subsection. The number density of the free quark system can be expressed as 
\begin{equation}\label{1t}
    n_i = \dfrac{g_i}{2\pi^2}\int_0^{\nu_i}p^2dp = \dfrac{g_i\nu_i ^3}{6\pi^2}
\end{equation}
and the energy density
\begin{equation}\label{2t}
    \varepsilon = \sum_i \dfrac{g_i}{2\pi^2}\int_0^{\nu_i}\sqrt{p^2 + m_i^2} p^2dp
\end{equation}
where the summation index $i$ runs over all the particles present {(i.e., $i = u,\;d,\;s,\;e({\rm electron}),\; \text{and}\; \mu ({\rm muon})$; representing the quarks and the leptons present in the system at this stage) and $g_i$ is the degeneracy factor}. Equations (\ref{1t}) and (\ref{2t}) are well-known expressions in literature. For free particles, $\nu_i$ and $\mu_i$ are related by the expressions
\begin{equation}
    \nu_i = \sqrt{\mu_i^2 - m_i^2} \qquad\text{or}\qquad \mu_i = \sqrt{\nu_i^2 + m_i^2}.
\end{equation}
It is general knowledge that quark masses depend on density and temperature. Ideally, the quark mass scaling is expected to be derived from the QCD theory, however, as it stands now, it is impossible due to the nonperturbative characteristics of the theory at low energies. In Ref.\cite{Peng:1999gh} a cubic root scaling law was reached at $T=0$ MeV based on an in-medium chiral condensate and was extended to a finite temperature treatment in Ref.\cite{Wen:2005uf}. Considering that the quarks interact with each other, we can introduce the interaction through density-dependent quark masses defined as 
\begin{equation}\label{3t}
    m_i = m_{i0} + m_I,
\end{equation}
where  $m_{i0}$ is the current quark mass and $m_I$ is the density dependent parameter representing quark interaction. We require that the relation for the number density from Eq.~(\ref{1t}) and the energy density from Eq.~(\ref{2t}) maintain their form after introducing Eq.~(\ref{3t}), which is the original idea of Ref.\cite{Fowler:1981rp}. Such masses have been named equivalent mass to differentiate them from other mass concepts \cite{Peng:2003jh}. Using the equivalent mass, both $\varepsilon$ and $n_i$ are required to maintain their original form like the free particle system, however, in the quasiparticle model as the one under consideration, the $n_i$ remains unchanged but the $\varepsilon$ changes. Such models are known in the literature as the quasiparticle models. A detailed derivation of the corresponding $\varepsilon$ in the quasiparticle model using fundamental thermodynamic consistency can be found in Sec. IIB of Ref.\cite{Peng:2008ta} and in Sec. II of Ref.\cite{Peng:2000ff} they used general ensemble theory. We do not intend to repeat these derivations here, however, the main point to note is that the Fermi momentum is not directly connected to the chemical potential, on the contrary, the chemical potential is replaced by an effective chemical potential $\mu_i^*$. Hence, the new form of $\nu_i$ becomes $\nu_i = \sqrt{\mu^{*2}_i - m^2_i}$, while the real chemical potential $\mu_i$ is related to the $\mu^*_i$ as
\begin{equation} \label{6t}
    \mu_i = \mu^*_i + \sum_j\dfrac{\partial \Omega_0}{\partial m_j}\dfrac{\partial m_j}{\partial n_i} \equiv \mu_i^* - \mu_I,
\end{equation}
then 
\begin{equation}
    n_i = \dfrac{g_i}{6\pi^2}\left(\mu^{*2}_i - m_i^2\right)^{3/2},
\end{equation}
with $\mu_I$ the interaction term. The thermodynamic potential $\Omega_0$ is determined to be
\begin{equation}
    \Omega_0(\{n_i\},\{m_i\}) = -\sum_i \dfrac{g_i}{24\pi^2}\left[\mu^*_i\nu_i\left(\nu^2_i-\dfrac{3}{2}m_i^2\right)+ \dfrac{3}{2}m_i^4\ln\dfrac{\mu_i^*+\nu_i}{m_i}\right],
\end{equation}
the pressure $P$ is expressed as 
\begin{align}\label{6at}
    P = -\Omega_0 + \sum_{i,j}n_i\dfrac{\partial \Omega_0}{\partial m_j}\dfrac{\partial m_j}{\partial n_i},
\end{align}
with 
\begin{equation}
    \dfrac{\partial\Omega_0}{\partial m_i}=\sum_i\dfrac{g_i m_i}{4\pi^2}\left[\mu_i^*\nu_i-m^2_i\ln\left(\dfrac{\mu_i^*+\nu_i}{m_i}\right)\right],
\end{equation}
and the energy density is  given by
\begin{equation}
    \varepsilon = \Omega_0 + \sum_i \mu^*_i n_i.
\end{equation}

We adopt the cubic root mass scaling law for the equivalent mass \cite{Peng:1999gh, Xia:2014zaa} expressed as 
\begin{equation}\label{m1}
    m_i = m_{i0} + \dfrac{D}{n_b^{1/3}} + Cn_b^{1/3} = m_{i0} + m_I,
\end{equation}
where $n_b$ is the baryon density and $D$ is a parameter associated with confinement, QCD string tension, chiral restoration density, and chiral condensate in a QCD vacuum. On the other hand, $C$ corresponds to the strength of one-gluon-exchange interaction, the first-order perturbative interaction, deconfinement phase transition, and it takes on positive values.

\subsection{Finite temperature formalism}\label{secdd2}

Since we are dealing with hot QM with masses dependent on temperature $T$, and baryon density $n_i$ of each particle $i$, it is convenient to work with state variables $T$, $n_i$, and the volume $V$. We start with the thermodynamic free energy {density} $F$ and derive the other quantities such as the energy density $\varepsilon$ and pressure $P$ through self-consistent thermodynamic treatment. We express the $F$ in the same way as the free particle system with the free particle mass and chemical potential replaced with $n_b$ and $T$ {dependent particle mass $m_i \rightarrow m_i(T,n_b)$ and chemical potential $\mu_i \rightarrow \mu^*_i(n_b, T)$}. Hence,
\begin{align}\label{1}
    F &= F(T,V,\{n_i\},\{m_i\})\nonumber\\
    &=\Omega_0(T, V,\{\mu^*_i\},\{m_i\}) + \sum_{i=u,d,s}\mu^*_in_i,
    \end{align}
where $\mu^*_i$ is the effective chemical potential of the $i$th particle and $\Omega_0$ is the thermodynamic potential of the noninteracting particles which should be considered as an intermediate variable for now. {The true thermodynamic potential $\Omega$ will be defined at the end of this section as a function of $\Omega_0$.} Since the independent state variables defined above do not include the $\mu^*_i$ that appears in $\Omega_0$, we need to find a mechanism to link it to the independent variables. We create such a linkage through the particle number density $n_i$ i.e. 
\begin{equation}\label{1a}
    n_i = -\dfrac{\partial}{\partial\mu^*_i}\Omega_0(T, V,\{\mu^*_i\},\{m_i\}),
\end{equation}
this expression has been used in previous thermodynamic treatments with $\mu^*_i$ replaced with the real chemical potential, $\mu_i$. Here, we use $\mu^*_i$ for thermodynamic consistency. Taking the derivative of Eq.~(\ref{1}) we have
\begin{equation}\label{2}
    dF = d\Omega_0 + \sum_in_id\mu^*_i + \sum_i\mu^*_idn_i,
\end{equation}
with 
\begin{equation}\label{3}
    d\Omega_0 = \dfrac{\partial\Omega_0}{\partial T}dT + \sum_i\dfrac{\partial\Omega_0}{\partial\mu^*_i}d\mu^*_i + \sum_i\dfrac{\partial\Omega_0}{\partial m_i}dm_i + \dfrac{\partial\Omega_0}{\partial V}dV,
\end{equation}
and 
\begin{equation}\label{4}
    dm_i = \dfrac{\partial m_i}{\partial T}dT + \sum_j\dfrac{\partial m_i}{\partial n_j}dn_j.
\end{equation}
Substituting Eqs.~(\ref{3}) and (\ref{4}) into Eq.~(\ref{2}) and  grouping the terms, we have
\begin{equation}
    dF = \left( \dfrac{\partial\Omega_0}{\partial T} + \sum_i\dfrac{\partial\Omega_0}{\partial m_i}\dfrac{\partial m_i}{\partial T}\right)dT + \sum_i\left(\mu_i^* + \sum_j\dfrac{\partial\Omega_0}{\partial m_j}\dfrac{\partial m_j}{\partial n_i}\right)dn_i + \dfrac{\partial \Omega_0}{\partial V}dV.
\end{equation}
Comparing the above expression with the thermodynamic relation
\begin{equation}
    dF = -SdT + \sum_i\mu_idn_i + \left(-P -F + \sum_i\mu_in_i\right)\dfrac{dV}{V},
\end{equation}
we can identify 
\begin{equation}\label{4a}
    S = -\dfrac{\partial \Omega_0}{\partial T} - \sum_i \dfrac{\partial m_i}{\partial T}\dfrac{\partial\Omega_0}{\partial m_i},
\end{equation}
where $S$ is the {entropy  density} of the particles, 
\begin{equation} \label{5}
    P = -F + \sum_i\mu_in_i - V\dfrac{\partial\Omega_0}{\partial V},
\end{equation}
and 
\begin{equation} \label{6}
    \mu_i = \mu^*_i + \sum_j\dfrac{\partial \Omega_0}{\partial m_j}\dfrac{\partial m_j}{\partial n_i} \equiv \mu_i^* - \mu_I.
\end{equation}
The last term in Eq.~(\ref{5}) persists when the finite size effect of the system cannot be neglected whether the particle masses are fixed or not. However, it can be ignored in this study just like other previous studies of DDQM \cite{Wen:2005uf, Peng:2000ff, Xia:2014zaa, Backes:2020fyw} considering an infinitely large system of QM like the one under consideration. In this case, the free energy density, Eq.~(\ref{1}), is independent of the volume of the system, thus the volume term in Eq.~(\ref{5}) does not appear. Now substituting Eqs.~(\ref{1}) and (\ref{6}) into Eq.~(\ref{5}) and considering an infinite QM system, the pressure becomes
\begin{align}\label{6a}
    P = -\Omega_0 + \sum_{i,j}n_i\dfrac{\partial \Omega_0}{\partial m_j}\dfrac{\partial m_j}{\partial n_i}.
\end{align}
The energy density is determined by substituting Eqs.~(\ref{1}) and (\ref{4a}) into the Helmholtz free energy density relation, $F = \varepsilon - TS$, yields
\begin{equation}\label{6b}
    \varepsilon = \Omega_0 + \sum_i\mu_i^*n_i - T\dfrac{\partial \Omega_0}{\partial T} - T\sum_i \dfrac{\partial m_i}{\partial T}\dfrac{\partial\Omega_0}{\partial m_i},
\end{equation}
and the actual thermodynamic potential becomes 
\begin{equation}
    \Omega = F - \sum_i\mu_in_i = \Omega_0 - \sum_{i,j}n_i\dfrac{\partial \Omega_0}{\partial m_j}\dfrac{\partial m_j}{\partial n_i}.
\end{equation}
Therefore, a given independent state variables $T$ and $n_i$, the $\mu_i^*$ can be determined by solving Eq.~(\ref{1a}) and the other thermodynamic quantities such as $S$, $\mu_i$, $P$, and $\varepsilon$ can also be calculated from Eqs.~(\ref{4a}), (\ref{6}), (\ref{6a}) and (\ref{6b}) respectively.

The net contribution of the $\Omega_0$ for the particles in the system can be expressed in terms of particles and anti-particles present, 
\begin{equation}
    \Omega_0^\pm = \Omega_0^+ + \Omega_0^-
\end{equation}
with $\Omega^+_0$ and $\Omega^-_0$ representing the contributions of the particle and anti-particle respectively. Explicitly,
\begin{equation}
    \Omega^\pm_0 = - \sum_i\dfrac{g_i T}{2\pi^2}\int_0^\infty p^2dp\left[\ln[1+e^{-(\epsilon_i-\mu_i^*)/T}] + \ln[1+e^{-(\epsilon_i+\mu_i^*)/T}]\right],
\end{equation}
where $g_i$ is the degeneracy factor and $i = u,\;d,\;s,\;e,\;\mu,\; \text{and}\; \nu_e$(electron neutrino) represents the quarks and leptons present in the system. The net number density, $n_i^\pm = n^+_i -n_i^-$ is given by
\begin{equation}
    n_i^{\pm}=-\dfrac{\partial \Omega_0^\pm}{\partial\mu^*_i} = \sum_i\dfrac{g_i}{2\pi^2}\int_0^\infty p^2dp\left[\dfrac{1}{1+e^{(\epsilon_i-\mu_i^*)/T}} - \dfrac{1}{1+e^{(\epsilon_i+\mu_i^*)/T}}\right],
\end{equation}
comprising particle ($n_i^+$) and antiparticle ($n_i^-$) respectively. The derivatives that appear in $S,\;\mu_i^*,\;P, \text{and},\;\varepsilon$ can also be determined as
\begin{equation}
    \dfrac{\partial\Omega_0^\pm}{\partial m_i} = \sum_i\dfrac{g_im_i}{2\pi^2}\int_0^\infty \left[\dfrac{1}{e^{(\epsilon_i-\mu^*_i)/T}+1}+\dfrac{1}{e^{(\epsilon_i+\mu^*_i)/T}+1}\right]\dfrac{p^2dp}{\epsilon_i},
\end{equation}
and 
\begin{align}
  \dfrac{\partial\Omega_0^\pm}{\partial T} = -\dfrac{g_i}{2\pi^2}\int_0^\infty\left(\ln\left[1+e^{-(\epsilon_i - \mu^*_i)/T}\right] + \dfrac{(\epsilon_i - \mu_i^*)/T}{1+e^{(\epsilon_i - \mu^*_i)/T}} + \ln\left[1+e^{-(\epsilon_i + \mu^*_i)/T}\right] + \dfrac{(\epsilon_i + \mu_i^*)/T}{1+e^{(\epsilon_i + \mu^*_i)/T}}\right)p^2dp,
\end{align}
where $\epsilon_i = \sqrt{(p^2+m_i^2)}$ is the single particle energy. 

The mass formula expressed in Eq.~(\ref{m1}) was extended to the finite temperature case \cite{Wen:2005uf, Chen:2021fdj} to study the deconfinement phase transition, in the form 
\begin{equation}\label{m2}
    m_i = m_{i0}+\dfrac{D}{n_b^{1/3}} \left(1 + \dfrac{8T}{\Lambda}e^{-\Lambda/T}\right)^{-1} + Cn_b^{1/3}\left(1 + \dfrac{8T}{\Lambda}e^{-\Lambda/T}\right),
\end{equation}
where $\Lambda = 280$ MeV is the QCD scale parameter whose value depends on the renormalization scheme \cite{Deur:2016tte, Peng:2006tk}, 
$C = 0.7$ (dimensionless) and $\sqrt{D}=130.6$ MeV. We use Eqs.~(\ref{m1}) and (\ref{m2}) as the equivalent mass relation with current quark masses for $u,\;d,\;{\rm and}\; s$ represented by $2.16$ MeV, $4.67$ MeV and $93.4$ MeV respectively as reported by the PDG \cite{Workman:2022ynf}.

\section{Properties of The SQM} \label{secprop}

The SQM under investigation is composed of $u,\;d,\;{\rm and}\; s$ quarks with leptons. In the neutrino-trapped matter phase (corresponding to the first and second stages of the star's evolution in the snapshots, Fig.~\ref{pfs}), we consider the three flavor quarks with electrons ($e$) and {their corresponding
neutrinos ($\nu_e$)}, since muons ($\mu$) only become relevant after the star has become neutrino-transparent in supernovae physics \cite{PhysRevC.100.015803}. In the neutrino transparent phase (the third and the last stages in the snapshots, Fig.~\ref{pfs}) matter consists of the three flavor quarks, $e$, and $\mu$. The quarks have degeneracy of $g_i = 6$ (3 colors $\times$ 2 spins), and the $e$ and $\mu$ with $g_i = 2$ while neutrinos are characterized with $g_i = 1$. The Fermi gas mixture then satisfies the $\beta$-equilibrium condition maintained by the weak interactions: $d,s\leftrightarrow u + e + \bar{\nu}_e$, $s+u\leftrightarrow u + d$, etc. In light of these interactions, the effective chemical potentials must satisfy the relations
\begin{equation}
    \mu^*_d = \mu_s^* =\mu_u^*+\mu_e -\mu_{\nu_e}, \qquad\text{neutrino trapped matter}\qquad
\end{equation}
and
\begin{equation}
    \mu^*_d = \mu_s^* = \mu_u^* + \mu_e, \qquad\text{neutrino transparent matter}\qquad
\end{equation}
with $\mu_e = \mu_\mu$ and $\mu_{\nu_\mu}=\mu_{\nu_e}$, the subscripts of the chemical potentials $\mu$ represent the type of particle under consideration. 
These expressions also hold for the real chemical potentials of the quarks as well.

The charge neutrality condition is given by the expression
\begin{equation}
    \dfrac{2}{3}n_u - \dfrac{1}{3}n_d-\dfrac{1}{3}n_s-n_e- n_\mu =0,
\end{equation}
the coefficients of the number densities ($n$) represent the 
electric charges of the particles and the subscripts represent the particle type, $n_\mu =0$ for the neutrino-trapped matter. The baryon number is conserved through
\begin{equation}
    n_b = \dfrac{1}{3}\left( n_u+n_d+n_s\right) = \dfrac{1}{3}\sum_in_i,
\end{equation}
and the {lepton fraction for the electron family} ($Y_{L,e}$) for the neutrino-trapped matter is conserved by
\begin{equation}
    Y_{L,e} = \dfrac{n_e+n_{\nu_e}}{n_b},
\end{equation}
where $Y_{L,e}=Y_e+Y_{\nu_e}$ and $Y_{L,\mu}=Y_\mu +Y_{\nu_\mu}=(n_\mu+n_{\nu_\mu})/n_b\approx 0$ according to supernovae physics \cite{PhysRevC.100.015803}. Also, we calculate the particle fractions $Y_i$ of each constituent quark in the system through the expression
\begin{equation}
    Y_i = \dfrac{n_i}{n_u+n_d+n_s},
\end{equation}
where $i$ runs over all the constituent quarks in the system.

\section{Results and discussions}\label{results}

\begin{table}[ht]
\begin{center}
\begin{tabular}{|c |c| c| c| c| c| c| c|}
\hline
  S$/$n$_B$ &Y$_{L,e}$ & M$_{max}[{\rm M}_{\odot}]$ & $R[{\rm km}]$ &$\varepsilon_0[{\rm MeV fm^{-3}}]$& $T_c[{\rm MeV}]$& $n_c[n_0]$& $M_{\rm b_ {max}}$\\
 \hline
1& $0.4$ & 2.20 & 13.86& 714.38& 8.82& 3.95 & 2.42\\  
2 & $0.2$ & 2.15 & 13.46&781.52 & 19.40 & 4.34& 2.42 \\
2 &$Y_{\nu e}= 0$ & 2.12& 13.40&749.867& 20.71 &4.28 & 2.43\\
\hline
$T = 0$ & $Y_{\nu e}= 0$ & 2.09& 13.24& 777.68& ---&4.49& 2.43\\
 \hline
 \multicolumn{8}{|c|}{Fixed $M_b$ and the corresponding stellar matter properties} \\
  \hline
   S$/$n$_B$ &Y$_{L,e}$ & $M_b[{\rm M}_{\odot}]$& M$[{\rm M}_{\odot}]$ & $R[{\rm km}]$&$\varepsilon_0[{\rm MeV fm^{-3}}]$&$n_c[n_0]$& $T_c[{\rm MeV}]$ \\
   \hline
   1& $0.4$ &\begin{tabular}[c]{@{}l@{}}$1.85$\\ $2.26$ \end{tabular}& \begin{tabular}[c]{@{}l@{}}$1.74$\\ $2.08$ \end{tabular}& \begin{tabular}[c]{@{}l@{}}$14.75$\\ $14.69$ \end{tabular}& \begin{tabular}[c]{@{}l@{}}$258.50$\\ $404.52$ \end{tabular}& \begin{tabular}[c]{@{}l@{}}{1.62}\\ {2.79} \end{tabular}& \begin{tabular}[c]{@{}l@{}}{5.97}\\ {7.61} \end{tabular}\\
   \hline
   2 & $0.2$ & \begin{tabular}[c]{@{}l@{}}$1.85$\\ $2.26$ \end{tabular} & \begin{tabular}[c]{@{}l@{}}$1.71$\\ $2.03$ \end{tabular}& \begin{tabular}[c]{@{}l@{}}$14.42$\\ $14.33$\end{tabular}& \begin{tabular}[c]{@{}l@{}}$276.26$\\ $424.25$ \end{tabular}& \begin{tabular}[c]{@{}l@{}}{1.80}\\ {2.62} \end{tabular}& \begin{tabular}[c]{@{}l@{}}{13.17}\\ {15.61} \end{tabular}\\
   \hline
   2 &$Y_{\nu e}= 0$ &\begin{tabular}[c]{@{}l@{}}$1.85$\\ $2.26$ \end{tabular} & \begin{tabular}[c]{@{}l@{}}$1.65$\\ $1.98$ \end{tabular}& \begin{tabular}[c]{@{}l@{}}$14.22$\\ $14.20$\end{tabular}& \begin{tabular}[c]{@{}l@{}}$274.28$\\ $416.36$\end{tabular}& \begin{tabular}[c]{@{}l@{}}{1.81}\\ {2.63} \end{tabular}& \begin{tabular}[c]{@{}l@{}}{14.06}\\ {16.70} \end{tabular}\\ 
   \hline
   $T = 0$ & $Y_{\nu e}= 0$ & \begin{tabular}[c]{@{}l@{}}$1.85$\\ $2.26$ \end{tabular}& \begin{tabular}[c]{@{}l@{}}$1.64$\\ $1.96$\end{tabular}& \begin{tabular}[c]{@{}l@{}}$14.11$\\ $14.06$\end{tabular}& \begin{tabular}[c]{@{}l@{}}$282.18$\\ $430.17$ \end{tabular}& \begin{tabular}[c]{@{}l@{}}{1.86}\\ {2.74}\end{tabular}& \begin{tabular}[c]{@{}l@{}}{---}\\ {---} \end{tabular}\\
  \hline
\end{tabular}
\caption{Stellar properties. Here, we show the maximum mass (M$_{\rm max}$) and its corresponding radii ($R$), the central energy density ($\varepsilon_0$), the central baryon density $n_c$, and the core temperature ($T_c$) for the evolution stages of the SQS considered. {Additionally, we choose two fixed baryonic masses; ${\rm M_b}=1.85\,{\rm M}_{\odot}$ and ${\rm M_b}=2.26\,{\rm M}_{\odot}$ and calculate the other properties of the stellar matter for the four stages of the stellar evolution.}} 
\label{T1}
\end{center}
\end{table}

\begin{figure}[ht]
  \includegraphics[scale=0.5]{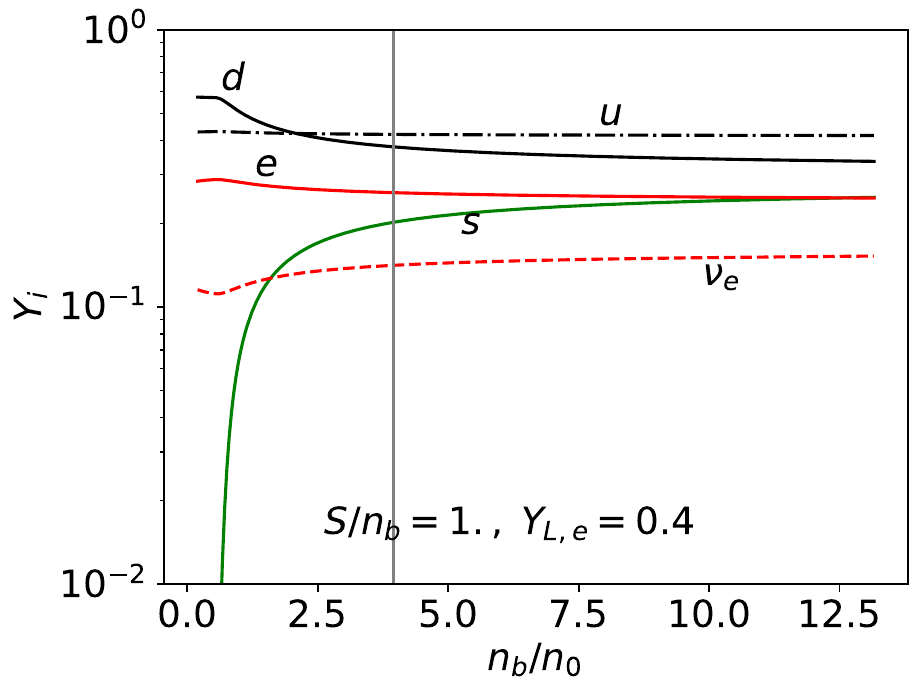}
  \quad
   \includegraphics[scale=0.5]{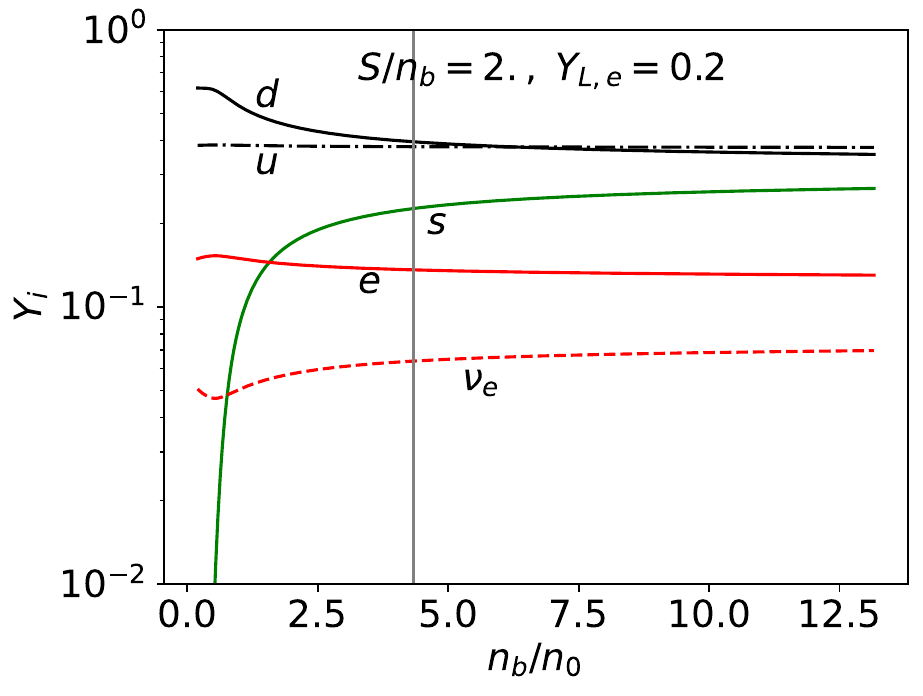}
  \quad
  \includegraphics[scale=0.5]{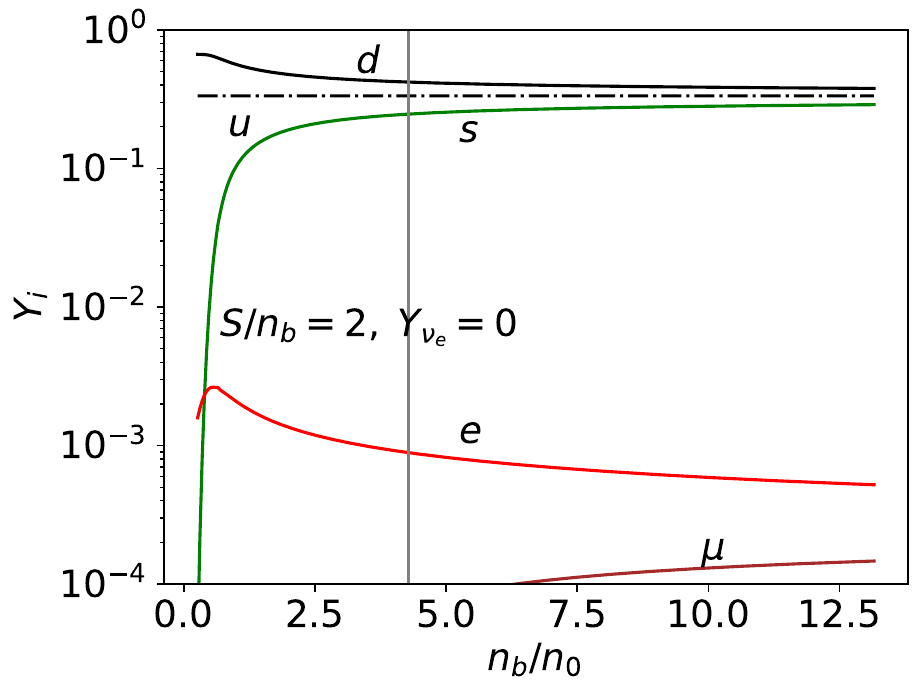}
  \quad
 \includegraphics[scale=0.5]{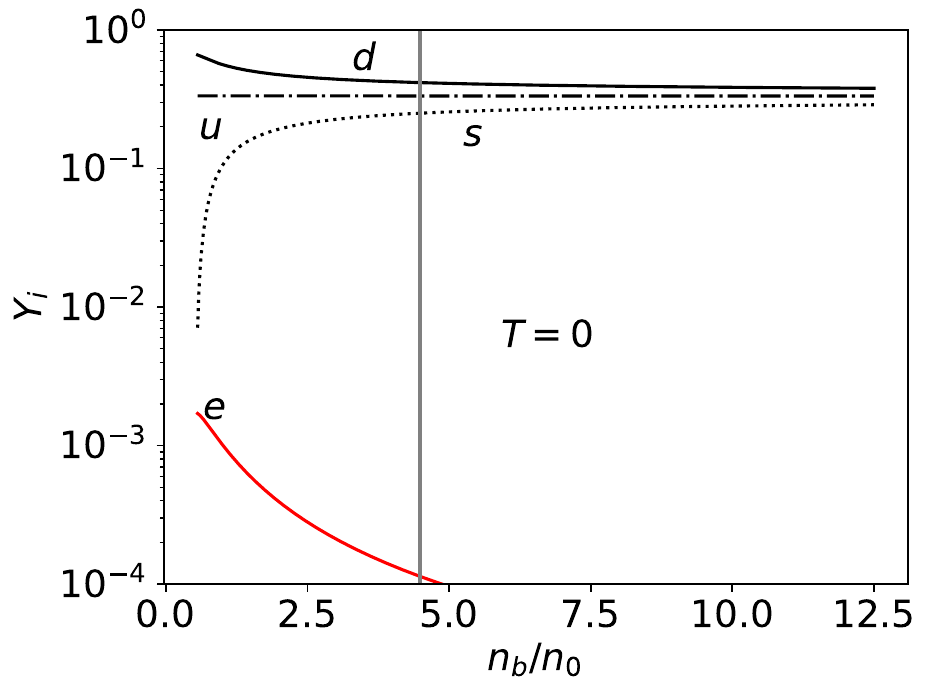}
\caption{In this diagram we show the snapshots of the particle distributions inside the star at various stages of its evolution. The upper panel shows the neutrino-trapped regime whilst the lower panel shows the neutrino-transparent regime of the star's evolution. The entropy densities $S/n_b$ are in units of Boltzmann constant $k_B$. The vertical gray line in the $Y_i$ represents the central baryon densities, the position at which the maximum and radii occur.}
    \label{pfs}
\end{figure}

In Fig.~\ref{pfs} we show the snapshots of the particle abundances, $Y_i$, in the evolution stages of the SQS from birth as PSS to maturity at $T=0$ MeV. {We used thermodynamic conditions that are characteristic of those of stellar collapse and core bounce \cite{Prakash:1996xs}. Generally, the initial condition before the eventual supernova explosion is denoted by $t\sim0$s in the literature.
This stage is represented by the first panel (top-left) characterized by $S/n_b \sim 1$ and $Y_{L,e} \sim 0.4$ as the initial stage of the star's evolution, similar studies have been done in detail in \cite{Prakash:2000jr, Janka:2006fh} using different classes of EoSs. For instance, in \cite{Burrows:1986me} the authors choose $S/n_b =0.8$ and $Y_{L,e} = 0.35$ to represent the initial condition of the star's evolution, and in \cite{Pons:1998mm, Raduta:2020fdn, Issifu:2023qyi, PhysRevC.100.015803, Sedrakian:2022kgj, Shao:2011nu} the authors choose $S/n_b =1$ and $Y_{L,e} = 0.4$ as their initial conditions, which is similar to the choice of this paper. Consequently, there is a closed set of thermodynamic conditions within which the evolution stages of PSSs can be chosen depending on the expected time interval under consideration.} In the second panel (top-right), the neutrinos have already started to diffuse, leading to a deleptonization ($Y_{L,e}=0.2$) and an increase in entropy density per baryon ($S/n_b =2$). The top panels (ambient condition of core birth) show the stages in which the star is neutrino-rich at the early stages of the star's birth. The first snapshot from the left represents the $Y_i$ of the newly born star before the supernova explosion, $t\sim 0{\rm s}$, and the second panel is when the star is about $0.5-1{\rm s}$ old after the core bounce. {Two general scenarios occur immediately after the supernova explosion, from the first to the second stage (top panels) of Fig.~\ref{pfs}: 1) If the supernova explosion is not strong enough to deleptonize the outer mantle of the PSS, the matter accretes again leading to the formation of a black hole. 2) If the supernova explosion is strong enough to generate a significant pressure loss in the surrounding disk through deleptonization, the mantle collapses and accretion becomes less important and the proto-star continues its evolution. This work assumes the second scenario because if the first one occurs, the evolution of the PSS terminates at the second stage, and it will not continue to the next stage.} 

The bottom panels (ambient conditions after deleptonization) represent the $Y_i$ when the star is neutrino-poor. The first snapshot at this stage represents when the star is about $10-15{\rm s}$ old, dominated by neutrino diffusion, deleptonization, and core heating. Here, the probability of the appearance of the strange quark increases relative to the previous stages. During the deleptonization and neutrino transparency period, the core continues to cool until the matter catalyzes when the entropy drops to $S/n_b =0$ at about $50{\rm s}$ later, the core then continues cooling through thermal emissions reaching $T = 0$ MeV in about $100$ years later \cite{Prakash:1996xs}.  {In between the second and the third stages (second-panel top and the first-panel bottom), when the star starts losing neutrinos rapidly, this can soften the EoS through deleptonization, this poses the possibility of black hole formation if the gravitational pressure is large enough. The evolution does not continue to the third stage if a black hole forms \cite{Prakash:1996xs}. After all the neutrinos have escaped from the core and the star is maximally heated (third stage in Fig.~\ref{pfs}), the probability of $s$ quark appearing in the stellar matter increases. This results in increasing the production of SQM, if the production of SQM is large enough, it will lead to the formation of a black hole \cite{Vidana:2002rg, Burgio:2006ed}. This phenomenon only occurs if there is accretion of matter at this stage. These processes have been described diagrammatically in \cite{Prakash:2000jr}. However, in this study, we assumed that the PSS evolved to its last stage of formation of cold-catalyzed SQS without forming a black hole along its evolution lines even though the possibility of black hole formation is not ruled out during the star's evolution.}

 Observing the snapshots across the top panels from left to right, we observe that the neutrino concentration reduces the relative abundance of $u$ and $d$ quarks and slightly suppresses the appearance of the $s$ quarks to higher $n_b$. Higher neutrino concentration means less $u$ and $d$ abundances and delayed appearance of the $s$ particle. A similar observation is made in the evolution of PNSs where both nucleons and exotic baryons are present in the stellar matter. In these studies, higher neutrino concentrations were observed to delay the appearance of particles with strangeness \cite{Issifu:2023qyi, Raduta:2020fdn, PhysRevC.100.015803, Sedrakian:2022kgj}. Also, the temperature rises in the core of the star as the entropy density per baryon increases and deleptonization of the stellar matter influences the appearance of the strange particle. The $s$ quark appears earlier in a hotter matter than relatively cold ones, as expected. The strange quark shifts towards the lower baryon density region from the first, second, and third stages of the star's evolution which reflects the temperature rise in the core of the star along its evolution lines presented in Tab.~\ref{T1}. In the last stage the appearance of the $s$ quark delays to a relatively higher baryon density. In the first stage, $s$ quark starts appearing at $n_b\sim 0.66n_0$, in the second stage, $s$ quark starts appearing at $n_b\sim 0.53n_0$, in the third stage it starts appearing at $n_b \sim 0.30n_0$ and in the final stage it starts appearing at $n_b\sim 0.59n_0$. Along the panels from top to bottom, left to right, the $Y_d$ and $Y_s$ increase while $Y_u$ decreases relatively, along the evolution lines of the star. It is worth mentioning that, in the lower panel, the muons do not appear in the $Y_i$ at $T=0$ because they constitute less than 0.01\% of the particle constituents. Also, the central baryon densities ($n_c$) are represented by vertical gray lines in Fig.~\ref{pfs}. It shifts towards the higher baryon density regions along the evolution lines of the star in the neutrino-trapped matter (top panels). In the neutrino transparent matter (bottom panels) the $n_c$ increases as well along the evolution lines. Similar behavior is observed with the central energy densities ($\varepsilon_0$) shown in Tab.~\ref{T1}. {From Tab.~\ref{T1}, we observe that the ${\rm M_{b_{max}}}$ remains constant for the neutrino trapped regime and that of the neutrino transparent matter also remains constant whilst the other properties of the stellar matter changes through the evolution stages. In the same table, we fixed two different ${\rm M_b}$ and determined the matter properties of these stars. The variation between the stellar radii for these stars along the evolution lines changes slightly but their gravitational masses change significantly. For a fixed M$_b$, both the gravitational masses and the radii decrease throughout the evolution stages of the stars, and the central energy density and the temperature increase as the stars evolve, before reaching `stability' at $T=0$.} It is worth mentioning here that the stiffness or otherwise of the EoS can be determined through the particle distribution in the stellar matter. It is known that the appearance of new degrees of freedom such as hyperons (strangeness-rich particles) {in hadronic stars} softens the EoS. A similar situation applies here, the higher the probability of the $s$ quark appearing in the stellar matter the softer the corresponding EoS and vice versa. Consequently, the relative maximum masses of the PSSs can be predicted through the particle distribution since stiffer EoS means a higher maximum mass.

From previous works by \cite{Chu:2012rd, DiToro:2006bkw, Shao:2012tu} the isospin asymmetry ($u-d$ quark asymmetry) was determined to be 
\begin{equation}
    \delta = 3\dfrac{n_d-n_u}{n_d+n_u},
\end{equation}
where $n_3 = n_d - n_u$ is the isospin density and $n_b= (n_d+n_u)/3$ is the baryon density for two flavor $u-d$ QM. We can deduce that for pure neutron (proton) nucleon matter $\delta =1(-1)$ based on their constituent quark structure. We observe along the evolution stages of the star that, the $Y_u$ decreases as $Y_d$ increases from the first stage of the star's evolution to maturity at the fourth stage. Hence, the stellar matter becomes more asymmetric as the star evolves.  The $u-d$ quark asymmetry decreases as $n_b$ increases towards the star's core. In the neutrino-trapped matter $\delta$ has a negative value at the higher density regions due to the higher abundance of $u$ quark relative to the $d$ quark beyond the $n_c$. We can infer that the star is proton-rich at this point of its evolution.

\begin{figure}[ht]
 \includegraphics[scale=0.7]{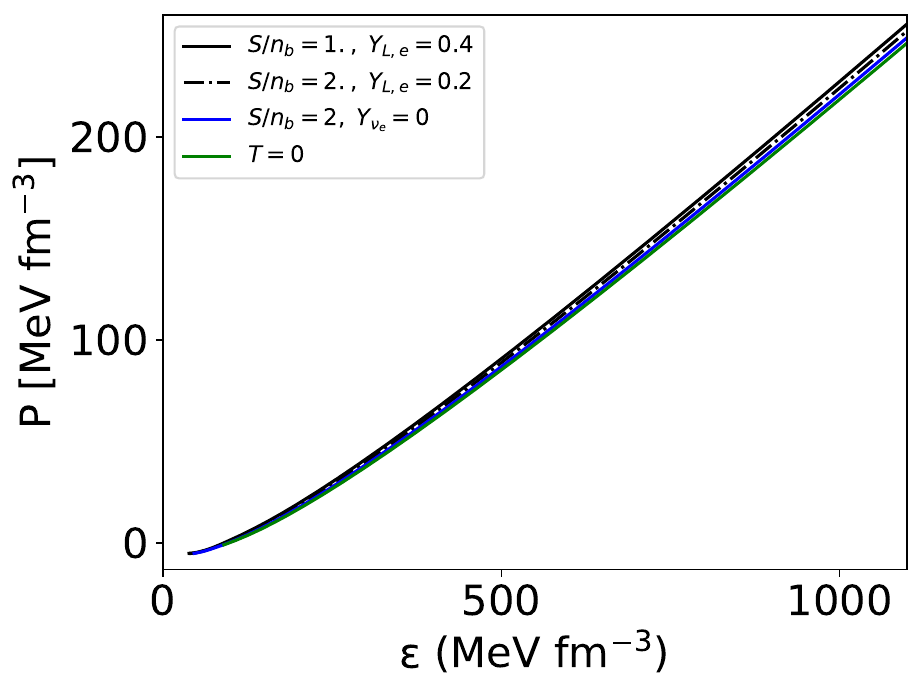}
\caption{The nuclear EoS composed of free quarks and leptons in $\beta$-equilibrium at different stages of stellar evolution.
}
    \label{eo}
\end{figure}

In Fig.~\ref{eo} we show, in units of MeV fm$^{-3}$, the EoS for the SQM at different stages of evolution of the SQS, calculated from the equations presented in Subsecs.~\ref{secdd1} and \ref{secdd2} and the required equilibrium conditions in Sec.~\ref{secprop}. We observe that the newly born star with higher neutrino concentration presents a stiffer EoS. As the star evolves through deleptonization its EoS starts to soften {as the probability of $s$ quark appearing in the stellar matter increases}. The EoSs become stiffer with the increase of the lepton number, mainly at high densities. Hence, the EoS softens from the first to the fourth stage as the star evolves. This reflects the reduction of the maximum mass of the star as it evolves from the first to the last stage as shown on Tab.~\ref{T1}. On the same table, we observe that fixed entropy stars are more massive than the colder ones, a similar observation was made in \cite{chu2021quark, Chu:2017huf}. 
From Fig.~\ref{pfs} we can deduce that lower lepton number concentration softens the EoS as we move along the panels.
Additionally, we can deduce that higher isospin asymmetry corresponds to softer EoS.

\begin{figure}[ht]  
 \includegraphics[scale=0.7]{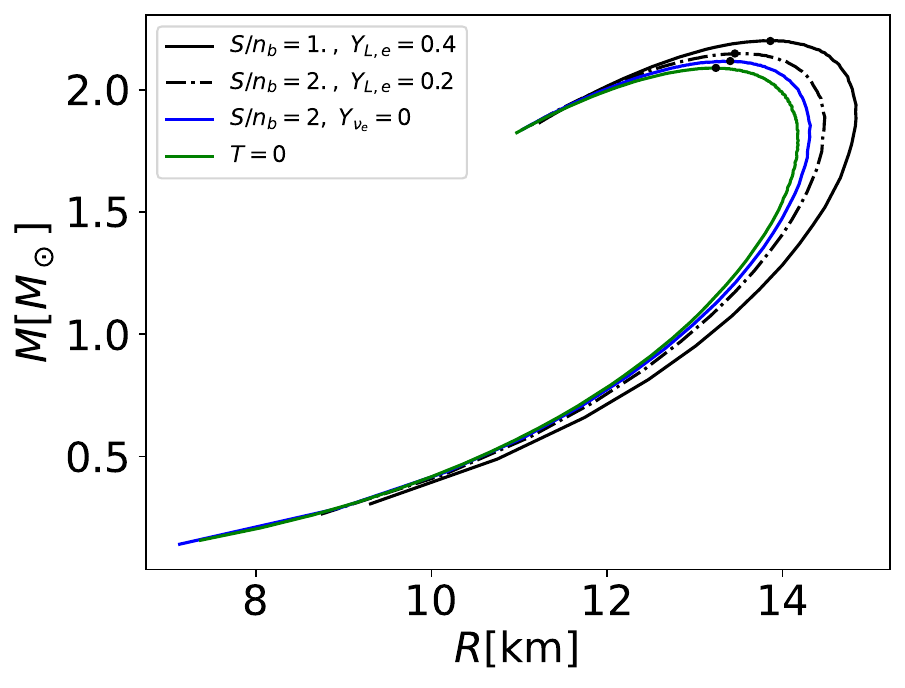}
\caption{The mass-radius relation for different stages in the evolution of SQS. 
The dark points on the curves represent the precise position of the star's maximum mass and the corresponding radii.}
    \label{mr}
\end{figure}

In Fig.~\ref{mr}, we determine the structure of the stars through their mass-radius (M-R) diagram. To find the M-R relation for the stars, we need to solve the Tolman–Oppenheimer–Volkoff (TOV) equations \cite{PhysRev.55.374} assuming static spherically symmetric fluid. The equations are expressed as
\begin{align}
    \frac{dP(r)}{dr}&=-[\varepsilon(r) + P(r)]\frac{M(r)+4\pi r^3 P(r)}{r^2-2M(r)r}, \label{eq1}\\
    \frac{dM(r)}{dr}&=4\pi r^2 \varepsilon(r), \label{eq2}
\end{align}
where $r$ is the radial coordinate, $M(r)$ is the gravitational mass, $P(r)$ is the pressure, and $\varepsilon(r)$ is the energy density, and we are using natural units ($G=c=1$). {
Our analysis 
is motivated by the quasi-static approximation for proto-neutron star evolution in the spherical symmetric form, which results in the consideration that the star is in hydrostatic equilibrium. This eliminates the time derivatives of density, pressure, and the metric, leading to the use of average hydrodynamic effects that concentrate on the evolution of intensive thermodynamic quantities such as the $Y_{L,l}$ and the $S/n_b$ over the Kelvin-Helmholtz timescale\cite{Pons:1998mm, roberts2012new}, unlike the full hydrostatic simulation approach where matter is not necessarily in $\beta$-equilibrium. Since we have a static spherically symmetric spacetime with Schwarzschild gauge, the TOV equations are applicable (see e.g. \cite{Raduta:2020fdn, Oertel:2016xsn, Sedrakian:2022kgj, PhysRevC.100.015803, Shao:2011nu, Chen:2021edy, Kumari:2021tik}) because the system is completely in hydrostatic equilibrium throughout the star's evolution. In this case, the boundary conditions are $P(R)=P_{\rm surf}$, with $P_{\rm surf}$ the surface pressure and $R$ is the radius of the star. 
In the specific case of a cold-catalyzed NS, $P_{\rm surf} = 0$. The choice of $P_{\rm surf}$ is arbitrary but has been found to have a significant effect on the characteristics of the mantle at the early stages of the star's evolution. On the other hand, it has been established that significantly low values of $P_{\rm surf}$ have a negligible impact on the evolution of the internal structures of the star over long timescales \cite{RevModPhys.74.1015, Gulminelli:2015csa}.}

In Fig.~\ref{mr}, we show the M-R relation for the different values of the $S/n_b$ considered in this work. The precise position of the maximum mass M$_{\rm max}$ and the corresponding $R$ for each curve is shown by a dark point. We observe that the maximum mass of the PSSs decreases throughout its time evolution. The highest values of M$_{\rm max}$ are achieved in the neutrino-trapped regime; in the first stage ($S/n_b = 1,\; Y_{L,e}=0.4$) with $M_{\rm max}=2.20$ M$_{\odot}$ and in the second stage ($S/n_b = 2,\; Y_{L,e}=0.2$) with $M_{\rm max}=2.15$ M$_{\odot}$. After all the neutrinos have escaped from the core of the star, its maximum mass is further reduced; in the third stage ($S/n_b = 2,\; Y_{\nu_e}=0$) we have $M_{\rm max}=2.12$ M$_{\odot}$ and the maximum mass continues to decrease until it reaches its smallest value, $M_{\rm max}=2.09$ M$_{\odot}$ when the star cools and becomes a stable SQS at the fourth stage. From Fig.~\ref{mr}, we can deduce that over time, the SQSs shrink and become smaller, so their radii decrease along the stages of evolution, achieving the smallest values when the star catalyzes and contracts to become a cold SQS. Comparing this result to the one obtained in Fig.~\ref{pfs}, we can deduce that lepton fraction and isospin asymmetry influence the mass of the star. A lower lepton fraction coupled with a high isospin asymmetry leads to a decrease in the maximum mass of the star and vice versa. As one always obtains,
a stiffer EoS corresponds to a higher maximum mass along the evolution lines of the SQS. It is important to mention that, the structure of the PSSs and the cold SQS studied here have maximum mass within the $2\,{\rm M_\odot}$ constraint required for NSs, such that, this model can accommodate the masses of PSR J0348+0432 ($2.01\pm 0.04\,\text{M}_\odot$), PSR J0740+6620 ($2.072_{-0.066}^{+0.067}$ M$_{\odot}$), and PSR J1614-2230 ($1.908 \pm 0.016$ M$_{\odot}$)\cite{demorest2010two,arzoumanian2018nanograv}, among others.
Besides that, the radii determined here are well in the range of the one determined for the massive millisecond pulsar PSR J0740+6620 mentioned in the introduction. 

\begin{figure}[ht]
 \includegraphics[scale=0.7]{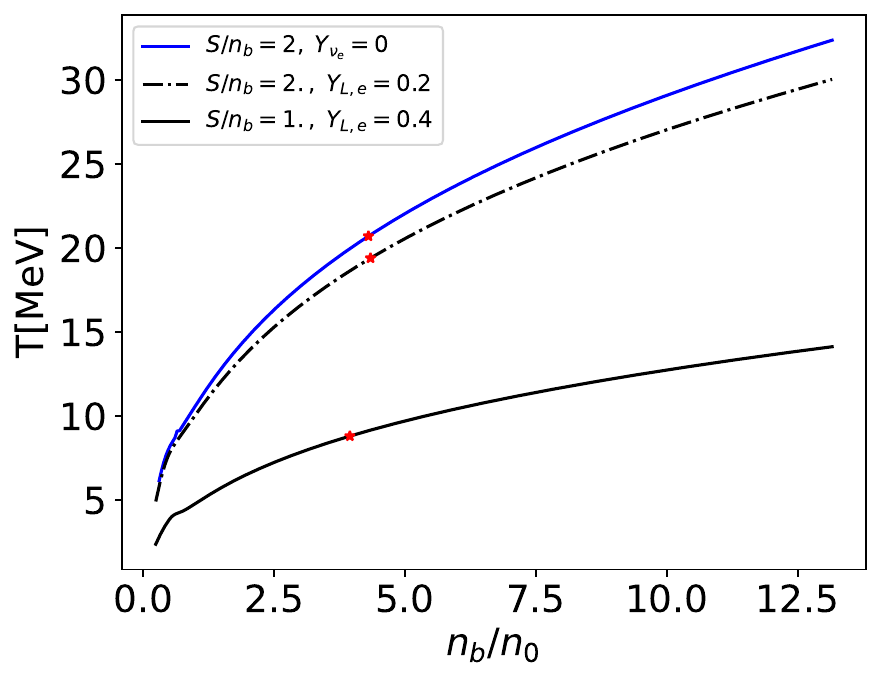}
\caption{The temperature variation as a function of the baryon number density for neutrino trapped and neutrino transparent stellar matter is presented in the left panel. 
The red stars on the curves represent the position of the core temperature of the stellar matter.}
    \label{figT}
\end{figure}

In {Fig. \ref{figT}}, we show the temperature $T$ variation in MeV as a function of $n_b/n_0$ for different values of the $S/n_b$. The core temperature, which corresponds to the central energy density of the stars where their maximum masses occur is shown with red stars on the plots with their corresponding $n_c$, see details on Tab.~\ref{T1}. In the first stage, when $S/n_b=1,\; Y_{L,e}=0.4$, the star is trapping neutrinos and heating up to cause an explosion, so it has a lower temperature profile. This stage has the lowest temperature profile compared to the other two stages studied that involve fixed entropies, with a core temperature of about 8.82 MeV at a central baryon density of $n_c=3.95n_0$. In the second stage, a few seconds after the core bounce, the trapped neutrinos start to diffuse, heating the stellar matter thereby increasing the entropy density ($S/n_b =2,\; Y_{L,e}=0.2$) and causing deleptonization as the neutrinos escape from the core of the star. This process heats the star, increasing its core temperature to about 19.40 MeV at a central baryon density of $n_c = 4.34n_0$. From this stage, the neutrinos continue to escape the core of the star until they have completely escaped the stellar matter, at this stage (third stage) the stellar matter is maximally heated with $S/n_b=2,\; Y_{\nu_e}=0$ which will start cooling through neutrino pair radiation and thermal emissions. This corresponds to a core temperature of 20.71 MeV at $n_c = 4.28n_0$. The second and the third stages of the star's evolution comprise the same $S/n_b=2$ with different neutrino concentrations but the temperature profiles of the stellar matter are significantly different. This is because the neutrino concentration suppresses the temperature of the stellar matter. Lastly, after becoming neutrino-transparent the star goes through catalyzation and continues to cool until it reaches $T=0$ MeV several years later. We can also observe that the central energy densities of the stellar matter increase from the first to the second stage when neutrinos are trapped, and then slightly decrease in the third stage when the neutrinos have escaped from the core of the star and rise again when the star reaches $T=0$ MeV. It is important to mention here that, the temperature profile for the proto-quark stars is significantly smaller \cite{chu2021quark, Bordbar:2020fqj, Cardoso:2017pbu} compared to its hadronic counterparts, PNSs \cite{Sedrakian:2022kgj, Issifu:2023qyi, Prakash:1996xs, Oertel:2016xsn, Raduta:2020fdn}. As a result, observing the temperature profile of a compact object can enable us to determine its composition with reasonable certainty.

\begin{figure}[ht]
 \includegraphics[scale=0.7]{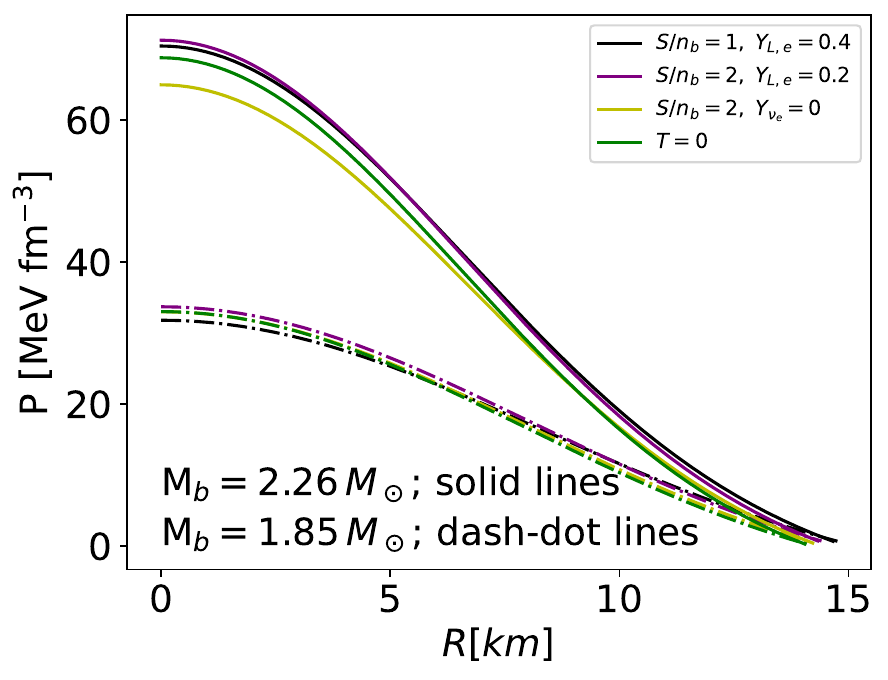}
\caption{Pressure profile of a PSS star with a fixed baryon mass. We considered two stars with fixed baryon masses and calculated their corresponding pressure and radii.}
    \label{pp}
\end{figure}
{In Fig.~\ref{pp}, we show the pressure variations in two stars with ${\rm M_b}= 1.85\,{\rm M_\odot}$ (represented by dot-dash lines on the graph) and ${\rm M_b}= 2.26\,{\rm M_\odot}$ (represented by solid lines on the graph) as a function of the stellar radii. We observe that the star with a larger ${\rm M_b}$ has a higher internal pressure and a higher M as shown on Tab.~\ref{T1}. Additionally, in the neutrino transparent regime, when $S/n_b = 2$ and $Y_{\nu_{e}}=0$, and the stellar matter is maximally heated the pressure in the star drops due to expansion, and at $T=0$ the star shrinks and its core pressure increases significantly.}

\begin{figure}[ht]
 \includegraphics[scale=0.7]{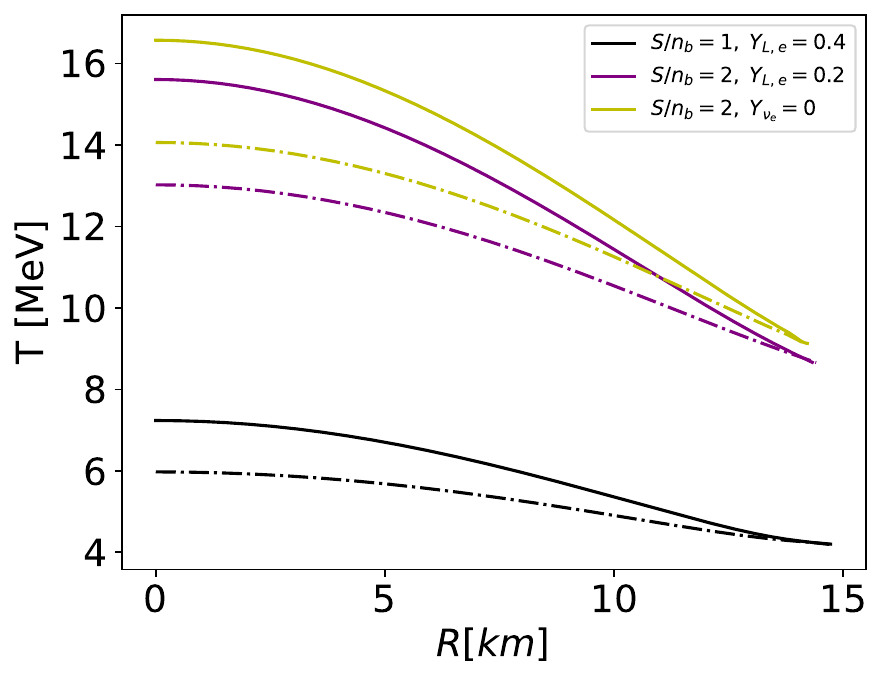}
\caption{The temperature profile of two stars with ${\rm M_b = 1.85\, M_\odot}$, represented by dashed-dot lines, and ${\rm M_b = 2.26\, M_\odot}$, represented by solid lines.}
    \label{TR}
\end{figure}
{Figure~\ref{TR} shows the temperature variations within the stars with  ${\rm M_b}= 1.85\,{\rm M_\odot}$ (represented by dot-dash lines on the graph) and ${\rm M_b}= 2.26\,{\rm M_\odot}$ (represented by solid lines on the graph) as a function of the stellar radii. As expected, the temperature of the stellar matter increases steadily towards the tar's core. Also, higher $S/n_b$ corresponds to a higher temperature, and when all the neutrinos have escaped from the star's core the temperature reaches its maximum as shown on the graph. We can also notice that, despite having different temperatures in their interiors, the two stars tend to attain the same surface temperature at each stage.}

\begin{figure}[ht]
  \includegraphics[scale=0.5]{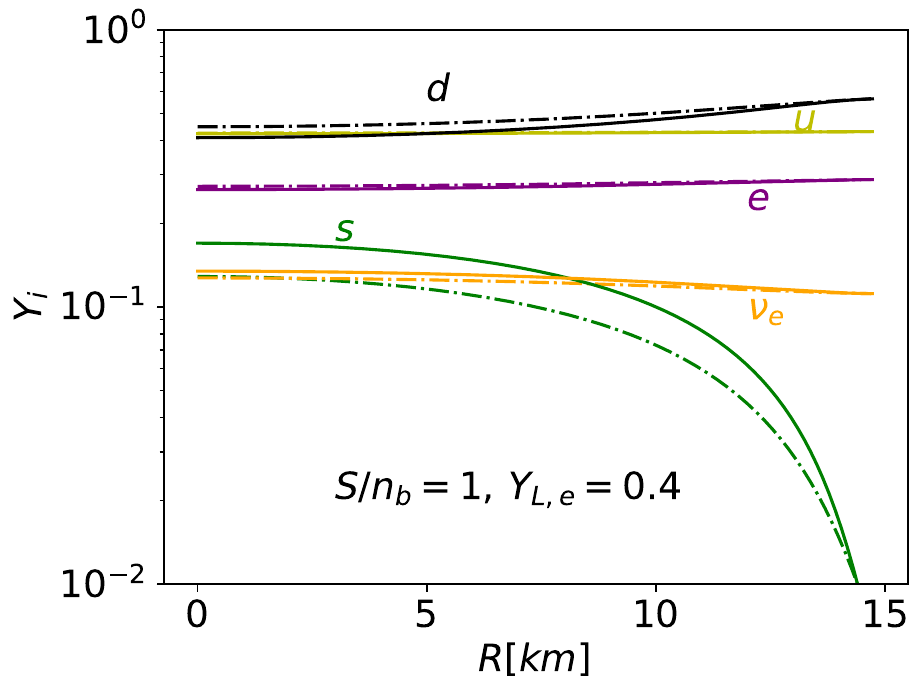}
  \quad
   \includegraphics[scale=0.5]{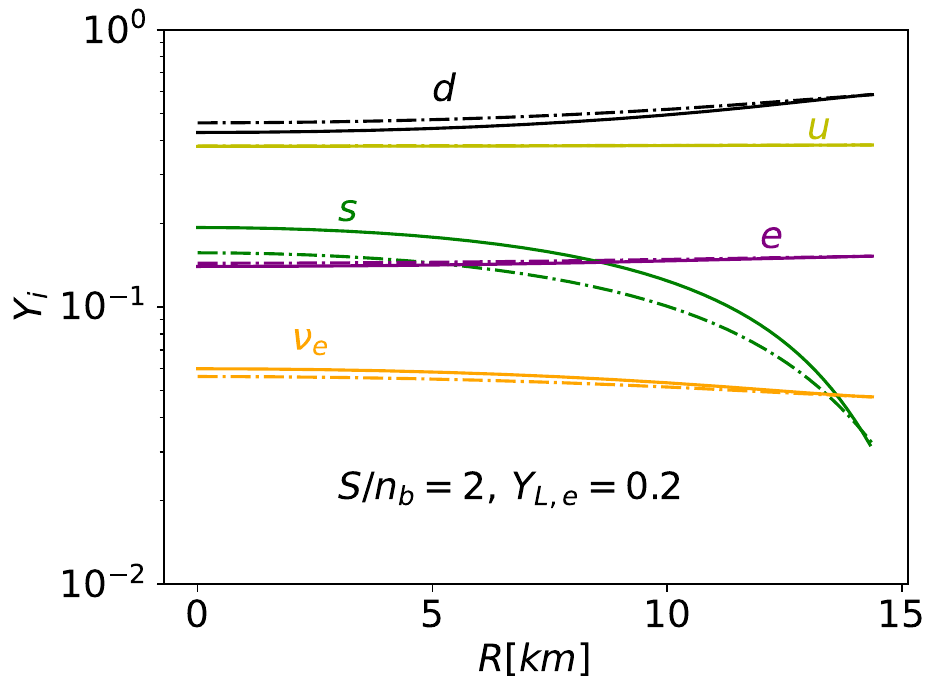}
  \quad
  \includegraphics[scale=0.5]{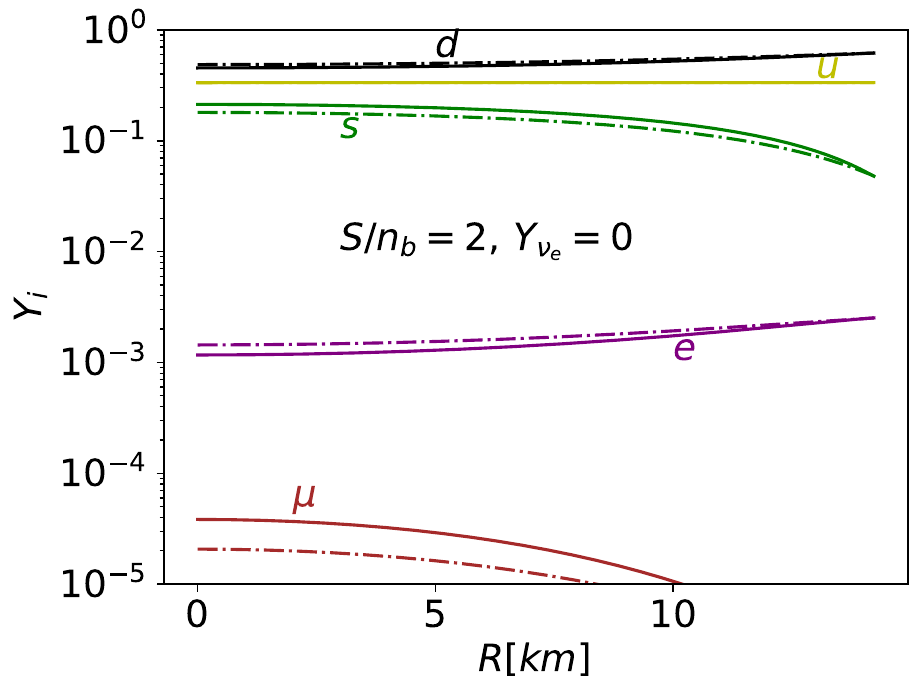}
  \quad
 \includegraphics[scale=0.5]{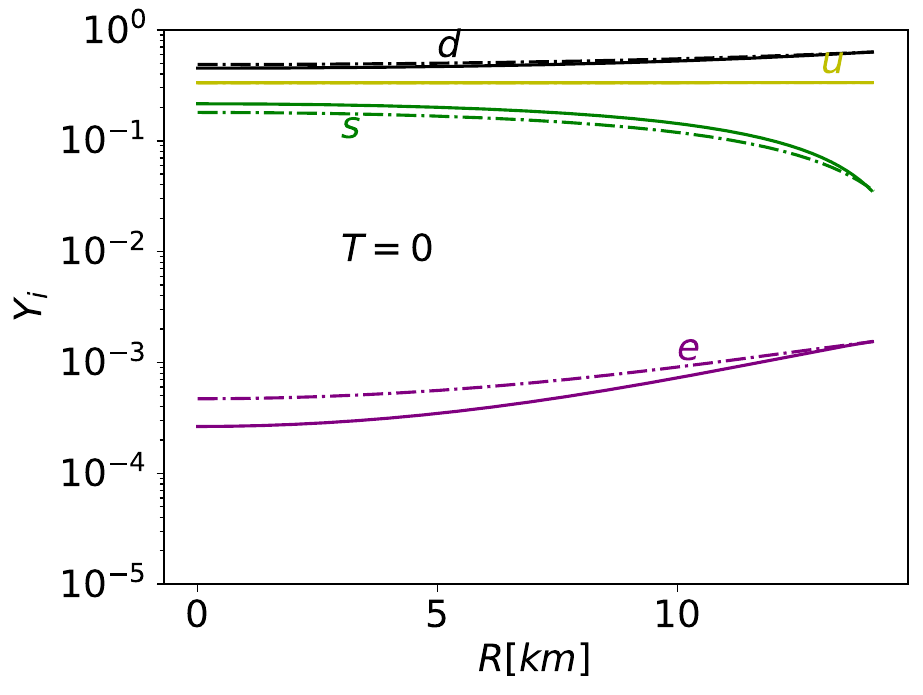}
\caption{Particle profile for two different stars. The dash-dot line represents ${\rm M_b = 1.85\, M_\odot}$ and the solid line represents ${\rm M_b = 2.26\, M_\odot}$. }
    \label{pfR}
\end{figure}
{In Fig.~\ref{pfR}, we present the particle profile as a function of the stellar radii. We show the particle profiles of two stars with ${\rm M_b}= 1.85\,{\rm M_\odot}$ (represented by dot-dash lines on the graph) and ${\rm M_b}= 2.26\,{\rm M_\odot}$ (represented by solid lines on the graph) on the same graph. We observe that, the particle population of the $d$ quarks and the $e$'s decrease slightly from the surface towards the core of stars, with the population of these particles for ${\rm M_b = 1.85\, M_\odot}$ slightly higher than that for ${\rm M_b = 2.26\, M_\odot}$ along the evolution lines. The $u$ quark remains fairly constant in both stars for the same thermodynamic conditions along the evolution lines, however, their population reduces from the first to the second stage and increases from the third to the final stage when the star is neutrino transparent. The $s$ quark and $\nu_e$ population increase from the surface towards the core along the evolution lines and, where $\mu$ are present, their appearance is delayed and rises from the center towards the core. This signifies that the production of $s,\, \nu_e$, and $\mu$ are largely temperature dependent. Moreover, their population increases with the size of the star.}

\begin{figure}[ht]
 \includegraphics[scale=0.7]{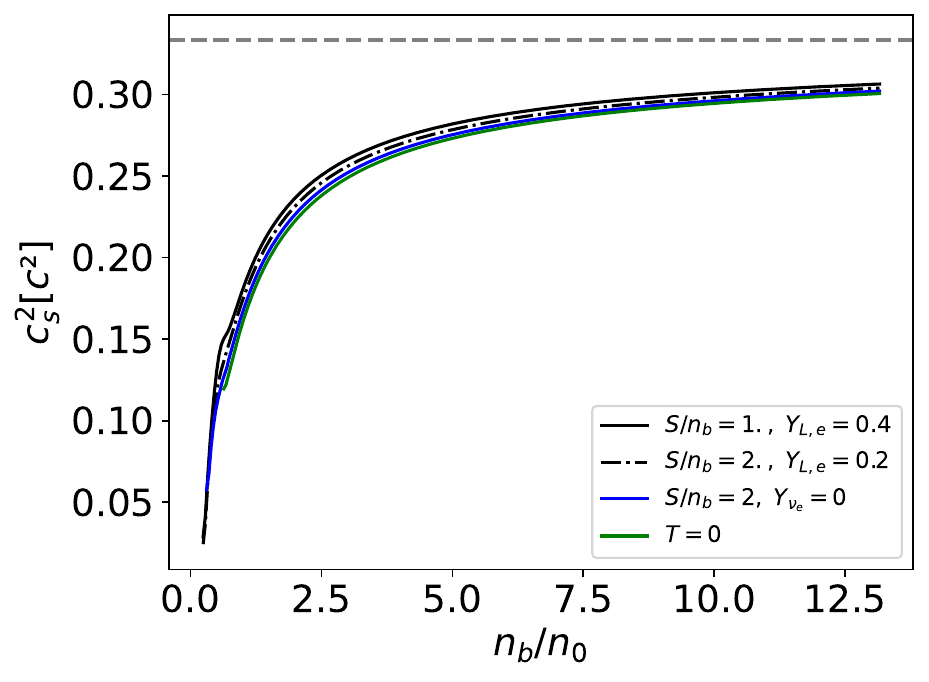}
\caption{The squared of sound speed as a function of $n_b/n_0$ for both neutrino rich and neutrino poor stellar matter. The horizontal gray line represents the conformal limit where $c_s^2 = 1/3$, below this limit matter is expected to behave in an asymptotically free manner, and above it, matter is expected to be in a confined state. }
    \label{sp}
\end{figure}

The strongly interacting matter under quantum chromodynamics (QCD) theory shows different properties at both the perturbative and the nonperturbative regions of the theory. These properties can be distinguished through conformal symmetry arguments. The nonperturbative matter phase (confined quark matter) is not symmetric under the conformal symmetry transformation due to chiral symmetry breaking. In contrast, the perturbative matter phase (deconfined quark matter) is approximately symmetric under the conformal symmetry transformation. These properties can be quantitatively distinguished through the study of the sound velocity $c_s$,
\begin{equation}
    c_s^2 = \frac{\partial P}{\partial \varepsilon},
\end{equation}
that uses the EoS as input. It is known that $c_s^2=1/3$ in exactly conformal matter while approaching the same value from below at high-density QM region, $n_b>40\,n_0$ \cite{Kurkela:2009gj}. In hadronic matter the $c_s^2$ can rise up to about $c_s^2\gtrsim 0.5$, causality requires that $c_s^2\leq 1$ and $c_s^2>0$ for thermodynamic stability \cite{Bedaque:2014sqa, Annala:2019puf}. In Fig.~\ref{sp}, we show the $c_s^2$ in units of $c^2$(constant speed of light) as a function of $n_b/n_0$ for the stages of the star's evolution. We observe that $c_s^2$ rises from the intermediate to the high baryon density regions as the star evolves and attains its lowest value at $T=0$ MeV. Therefore, the $c_s^2$ in hotter stars is higher than the ones in relatively colder stars. Also, the $c_s^2$ monotonically increases with increasing $n_b/n_0$ and approaches $c_s^2=1/3$ from below at high-density regions, for all the stages considered, which is in agreement with the results obtained for conformal QCD matter \cite{fraga2014interacting,borsanyi2010qcd, cherman2009bound}. Thus, our results show that in the case of the evolution of SQSs within the framework of DDQM, it is possible to have compact objects with $M_{\rm max} \geq 2$ M$_\odot$ formed by self-bound free quarks that conform with the conformal symmetry constraint for the speed of sound. 

\begin{figure}[ht]
 \includegraphics[scale=0.7]{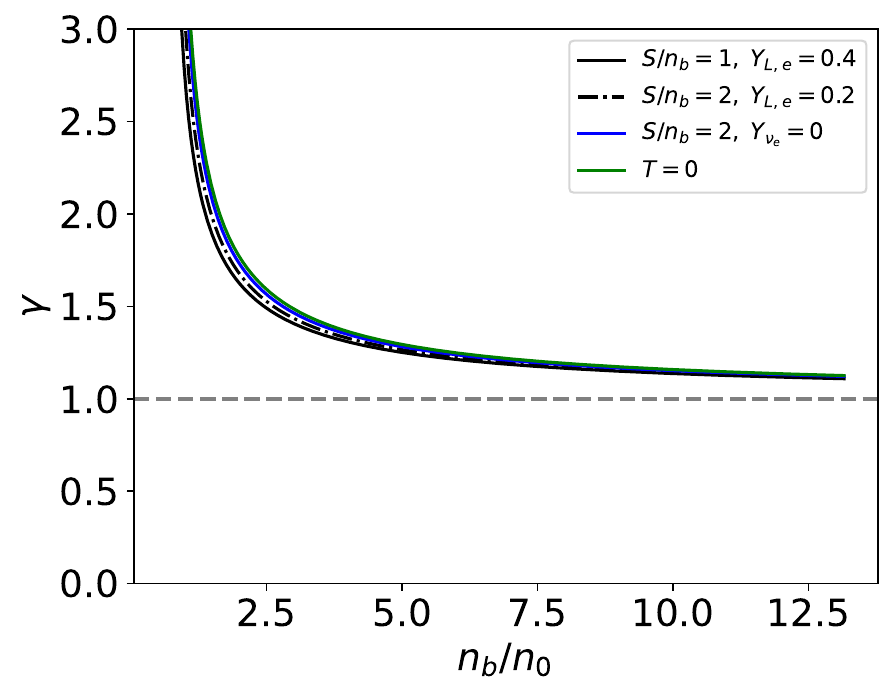}
\caption{The polytropic index as a function of $n_b/n_0$. The horizontal gray line, $\gamma =1$, represents the conformal boundary. Below this limit, QM is considered to be in a deconfine state, and above it, QM is considered to be in a confined state.}
    \label{pt}
\end{figure}

The polytropic index $\gamma$ is also used as a criterion for determining the stellar matter's inner composition. The $\gamma$ is mathematically expressed as 
\begin{equation}
    \gamma = \frac{\partial \ln P}{\partial \ln \varepsilon},
\end{equation}
and uses the EoS as its input. For matter with exact conformal symmetry, matter without intrinsic scales, $\gamma = 1$, independent of any coupling strength. At this symmetry when matter is independent of any dimensionful parameter, energy density and pressure become proportional to each other which leads to $\gamma = 1$. This symmetry has also been proven to yield $c_s^2 = 1/3$ as discussed above. On the other hand, chiral effective theory calculations and hadronic models predict $\gamma\approx 2.5$ around and above nuclear saturation density \cite{Kurkela:2009gj, Liu:2023gmq, Annala:2019puf}. This is because at low to intermediate densities of QCD matter, the ground state does not possess the approximate chiral symmetry of the Lagrangian that governs it \cite{Holt:2009ty}. The spontaneously broken symmetry results in the emergence of mass scales, like hadrons in the nuclear matter and scale-dependent interactions. In Fig. \ref{pt}, we show $\gamma$ as a function of $n_b/n_0$, the graph shows a steady decrease of $\gamma$ with $n_b/n_0$, which slowly approaches the conformal limit at the high-density region when $\gamma$ reaches a minimum value of $1.12$. The $\gamma\approx 1.12$ value is in good agreement with conformal matter classification considering that the limit set in Ref.\cite{Annala:2019puf} after examining a wide collection of EoSs is 1.75. The authors differentiated between hadronic and quarkionic matter phases with $\gamma \leq 1.75$, which is the average between the $\gamma$ calculated from perturbative QCD and chiral effective theory. 

\section{Conclusion}\label{conc}

In this work, we investigate the evolution of SQSs from birth as PSSs to maturity when the star cools down to $T=0$ MeV using a density-dependent quark mass model. The study assumes that the entire proto-star before the supernova explosion and after the core bounce is formed entirely by a deconfined SQM, so there are no phase transitions involved, PSSs are present from the beginning. We intend to extend the study to cover hybrid NSs in the future where PNS to PSS phase transitions become relevant \cite{Shao:2011nu}. We considered four different stages of the star's evolution: Two stages when neutrinos are trapped in the core of the star and the other two stages when the neutrinos have escaped from the core. The neutrino-trapped stellar matter was investigated by fixing the entropy density per baryon (stage 1: $S/n_b=1$; stage 2: $S/n_b=2$) and the lepton fraction (stage 1: $Y_{L,e}=0.4$; stage 2: $Y_{L,e}=0.2$), we found that the EoS is stiffer when the entropy density per baryon of the stellar matter is low, and the lepton fraction is relatively high. In this phase, the stellar matter is less asymmetric, and the central energy and baryon densities increase from low entropy density per baryon to higher entropy density per baryon accompanied by an increase in core temperature from 8.82 to 19.40 MeV {for stars with $\rm M_{max}$, see Tab.~\ref{T1}. In the same table, we show the results for the core temperatures for two different stars with fixed $\rm M_b$, their values are significantly lower than the ones obtained for the stars with $\rm M_{max}$.} In the neutrino-transparent phase, we consider the case where $S/n_b =2,\; Y_{\nu_e}=0$ (stage 3) and $T=0$ MeV (stage 4). In these phases, we found that the EoS for hot matter is stiffer than the one for cold SQM. The stellar matter becomes relatively more asymmetric compared to the previous stages and the central energy and baryon densities also increase from $S/n_b=2$ to $T=0$ MeV. The star has a higher core temperature of $T_c = 20.71$ MeV for $\rm M_{max}$ when all the neutrinos have escaped and $S/n_b =2$, {for stars with a fixed $\rm M_{b}$ the higher $T_c$ also occurs at this stage,} these results have been elaborated on Tab.~\ref{T1}. The stages of NS evolution from supernova explosion to maturity with the necessary thermodynamic conditions have been detailed in Refs.\cite{Janka:2006fh, Prakash:1996xs}. 

For the first time, we use DDQM to study the evolution of PSSs from birth, through deleptonization to the formation of cold-catalyzed SQS in a procedure similar to the one used in investigating PNSs in the literature \cite{Issifu:2023ovi, Oertel:2016xsn, Raduta:2020fdn}. That notwithstanding, some progress has been made by several authors in describing the structure of the PSSs using different phenomenological quark models adopting different approaches. For instance, {in \cite{Chen_2022}, the authors} explore strange quark matter and the structure of PSS using the same model as the one used in this work with a focus on conditions characteristic of the early formation of the star (e.g. they considered $S/n_b = 1$ and 0.5 with $Y_{L,e}=0.4$). Their treatment did not take into account the deleptonization stages. In \cite{Kumari:2021tik} the authors adopted the treatment of PSS similar to ours using the Polyakov chiral quark mean-field model considering conditions comparable to the birth of the star ($S/n_b=1$; $Y_{L,e}=0.4$) and neutrino transparent ($S/n_b=2$; $Y_{\nu,e}=0$) stages. Their results qualitatively agree with ours in those instances.

Additionally, we investigated the particle abundances in the core of the star at different stages of the star's evolution and associated it with other properties of the SQM in Fig.~\ref{pfs}. We observed that the isospin asymmetry of the stellar matter increases as the star evolves from the first to the fourth stage. Similar observations were made in Ref.\cite{Issifu:2023qyi} when they studied PNS with exotic baryons in its core.  We also studied the EoS of the SQM and presented our findings in Fig.~\ref{eo}, we observed that the EoS softens as the star evolves from the first to the fourth stage which coincides with the direction of increasing isospin asymmetry. Some discussions on isospin asymmetry and its effect on the evolution of SQSs can be found in Refs.\cite{chu2021quark, Chu:2017huf}. We determined the structure of the star by calculating its M-R diagram in Fig.~\ref{mr}, we observed that both the mass and the radii decrease as deleptonization occurs and the star evolves from the first to the fourth stage. The PSSs down to the stable SQS satisfy the $2\,{\rm M_\odot}$ constraint imposed by the observed pulsars PSR J0348+0432, PSR J0740+6620, and PSR J1614-2230 and the radii also satisfies the constraint imposed by the pulsar PSR J0740+6620 observed by NICER. We investigated the temperature profile in the star's core for the first three stages where entropy per baryon in the stellar matter was fixed and presented our outcome in Fig.~\ref{figT}. Generally, the temperature profile rises steadily with $n_b/n_0$. The first stage is when the star traps neutrinos, heats up and expands to cause an explosion, here, the stellar matter has the minimum temperature profile. The second stage is when the deleptonization starts with neutrinos escaping from the core, the stellar matter has the intermediate temperature profile. In the third stage when all the neutrinos have escaped from the core of the star, here, the stellar matter has its maximum temperature profile just before it starts cooling. This follows previous studies of PSSs \cite{chu2021quark, Cardoso:2017pbu, Bordbar:2020fqj} and PNSs \cite{Issifu:2023ovi, Oertel:2016xsn, Raduta:2020fdn}. {We also analyzed the pressure profiles of stars with a fixed $\rm M_{b}$ in Fig. \ref{pp}, where we observed that stars with higher baryonic masses have higher interior pressures. Also, the pressure inside these stars drops immediately after the neutrinos have escaped from their core, and then increases again at the last stage when the stars cool and shrink. The temperature profiles for the stars with fixed $\rm M_{b}$ were also studied, and the results shown in Fig. \ref{TR}, the temperatures for these stars are smaller at their surface and increase towards their center. Lastly, we also analyzed the particle fractions in the four stages of the star's evolution with the fixed $\rm M_{b}$'s in Fig. \ref{pfR}. We observed that the $d$ quark and the $e$'s decrease from the surface towards the star's core, the $u$ quark remains fairly constant for both baryon masses and $s$ quark, $\nu_e$ and $\mu$ population increase from the surface towards the core in the direction of increasing temperature. Generally, some analysis for fixed baryon masses can be found in \cite{Nakazato:2020ogl, Martinon:2014uua} and the references therein}.

Furthermore, we investigated the behavior of the QM that are assumed to form the stars by studying the sound velocity and presented the result in Fig.~\ref{sp}. The sound velocity imposes a clear conformal boundary on the QM determined from the EoS, which helps to differentiate between the hadronic and quarkionic matter phases. The stars determined here produce $c_s^2$ that lies below the $c_s^2=1/3$ conformal line. This is evidence that the stars investigated are formed by self-bound free quark matter. Finally, we studied the polytropic index and presented our result in Fig.~\ref{pt}. The polytropic index also helps classify matter phases using conformal symmetry arguments similar to $c_s$, here our results lie slightly above the conformal line, $\gamma \approx 1.12$. However, this agrees with conformal matter classification following the new limit $\gamma\leq 1.75$ set by the authors in \cite{Annala:2019puf}.

\section{Acknowledgements}

This work is a part of the project INCT-FNA Proc. No. 464898/2014-5. D.P.M. was partially supported by Conselho Nacional de Desenvolvimento Científico e Tecnológico (CNPq/Brazil) under grant 303490-2021-7.  A.I. was supported financially by Fundação de Amparo à Pesquisa do Estado de São Paulo (FAPESP) under grant number 2023/09545-1. F.M.S. would like to thank FAPESC/CNPq for financial support under grant 150721/2023-4.

\bibliographystyle{ieeetr}
\bibliography{references.bib}

\end{document}